\documentclass[
 aip,
 jmp,
 amsmath,amssymb,
reprint,
]{revtex4-1}

\newcommand{\sout}[1]{}

\usepackage{graphicx}
\usepackage{dcolumn}
\usepackage{bm}
\usepackage{color}

\begin{document}

\preprint{AIP/123-QED}

\title{A compact design for the Josephson mixer: the lumped element circuit}

\author{J.-D. Pillet}
\affiliation{Laboratoire Pierre Aigrain, Ecole Normale Sup\'erieure-PSL Research University, CNRS, Universit\'e Pierre et Marie Curie-Sorbonne Universit\'es, Universit\'e Paris Diderot-Sorbonne Paris Cit\'e, 24 rue Lhomond, 75231 Paris Cedex 05, France}
\affiliation{Coll\`ege de France, 11 place Marcelin Berthelot, 75005 Paris, France}

\author{E. Flurin}

\affiliation{Laboratoire Pierre Aigrain, Ecole Normale Sup\'erieure-PSL Research University, CNRS, Universit\'e Pierre et Marie Curie-Sorbonne Universit\'es, Universit\'e Paris Diderot-Sorbonne Paris Cit\'e, 24 rue Lhomond, 75231 Paris Cedex 05, France}

\author{F. Mallet}
\email[corresponding author: ]{francois.mallet@lpa.ens.fr}
\affiliation{Laboratoire Pierre Aigrain, Ecole Normale Sup\'erieure-PSL Research University, CNRS, Universit\'e Pierre et Marie Curie-Sorbonne Universit\'es, Universit\'e Paris Diderot-Sorbonne Paris Cit\'e, 24 rue Lhomond, 75231 Paris Cedex 05, France}

\author{B. Huard}

\affiliation{Laboratoire Pierre Aigrain, Ecole Normale Sup\'erieure-PSL Research University, CNRS, Universit\'e Pierre et Marie Curie-Sorbonne Universit\'es, Universit\'e Paris Diderot-Sorbonne Paris Cit\'e, 24 rue Lhomond, 75231 Paris Cedex 05, France}

\date{\today}
\begin{abstract}
We present a compact and efficient design in terms of gain, bandwidth and dynamical range for the Josephson mixer, the superconducting circuit performing three-wave mixing at microwave frequencies. In an all lumped-element based circuit with galvanically coupled ports, we demonstrate non degenerate amplification for microwave signals over a bandwidth up to 50 MHz for a power gain of 20 dB. The quantum efficiency of the mixer is shown to be about 70$\%$ and its saturation power reaches $-112$~dBm.
\end{abstract}

\maketitle

Analog processing of microwave signals has recently entered the quantum regime owing to the developments of superconducting circuits. Quantum limited amplifiers that are based on the non-linearity provided by Josephson junctions have been developed in various designs~\cite{Manuel2007, Tholen2007, Yamamoto2008, Kinion2008, Manuel2008, Kamal2009, Spietz2010, Eichler2011, Vijay2011, Hatridge2011, Gao2011, Hover2012, Eom2012, Bockstiegel2014, Mutus2013, Zhong2013, Mutus2014,Eichler2014,Narla2014,Zhou2014}.
Non-degenerate three-wave mixing, a key operation, is realized by the Josephson ring modulator (JRM), which is a ring of four identical Josephson junctions~\cite{Bergeal2010,Abdo2013c}. This element is at the core of several tools able to generate and manipulate quantum microwave modes such as phase preserving amplifiers~\cite{Bergeal2010b,Abdo2011,Roch2012},  non-local entanglement generators~\cite{Flurin2012}, frequency converters~\cite{Abdo2013}, quantum memories~\cite{Flurin2013} or circulators~\cite{Abdo2013b,Sliwa2015}. In all previous implementations of the Josephson mixer, the JRM was embedded at the crossing of two distributed or lumped resonators, which puts a constraint on bandwidth and dynamical range that can be detrimental to quantum operations. In this letter, we discuss the origin of this constraint and how to optimize the figures of merit of the Josephson mixer. These ideas are put in practice on an experiment in which a phase preserving amplifier is solely built out of a JRM that is shunted with lumped plate capacitors. Compared to previous implementations, we report an order of magnitude increase of its dynamical bandwidth, up to 50~MHz at a power gain of 20~dB while keeping the dynamical range as high as $-112~\mathrm{dBm}$.

\begin{figure}
\begin{centering}
\includegraphics[scale=.5]{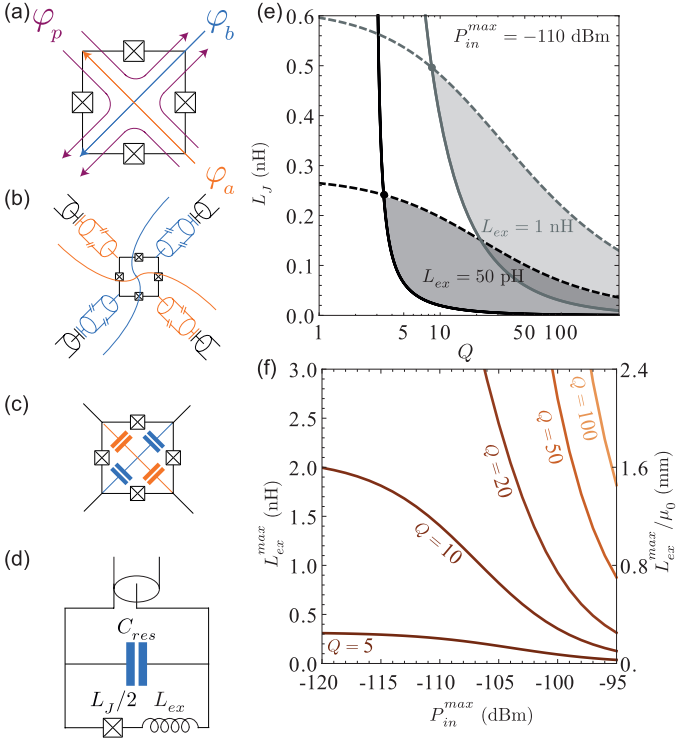}
\caption{\label{fig1} \textbf{(a)} Flux quadratures of the eigenmodes of the Josephson Ring Modulator. \textbf{(b)} Original mixer design where the JRM is placed at the crossing of two resonators of different frequencies. Two lines represent the spatial profiles of the voltage in the $\lambda/2$ resonators. \textbf{(c)} Ideal design where every inductive parts originate from the Josephson junctions. \textbf{(d)} Equivalent circuit of the differential modes $a$ and $b$ when the pump is turned off. A parasitic inductance $L_{ex}$ in series with the JRM is considered. \textbf{(e)} The two shaded regions, corresponding to $L_{ex}=1~$nH and $L_{ex}=50~$pH, indicate the combinations of quality factor $Q$ and Josephson junction inductance $L_J$ that verify the inequalities (\ref{pqeq}) and (\ref{qpeq}) for a power gain $G=20~$dB, a saturating input power $P_{in}^\mathrm{max}=-110~$dBm and operating frequency $f=7~$GHz. For a given extra inductance $L_{ex}$ and saturating power $P_{in}^\mathrm{max}$, there is a minimum quality factor which limits the bandwidth of the Josephson mixer and a maximum inductance  for the Josephson junction.   \textbf{f)} Maximal allowed value of the resonator extra inductance $L_{ex}^\mathrm{max}$ within the constraints (\ref{pqeq}) and (\ref{qpeq}) as a function of saturating power $P_{in}^\mathrm{max}$ plotted for various quality factors $Q$.}
\end{centering}
\end{figure}

In a Josephson mixer, the JRM couples three independent fluxes
 $\varphi_a$, $\varphi_b$ and $\varphi_p$ (Fig.~\ref{fig1}(a)), through the lowest order coupling term $H_{mix}=-E_J\sin(\Phi/4\varphi_0)\varphi_a\varphi_b\varphi_p$, where $E_J$ is the Josephson energy of each junction, $\varphi_0=\hbar/2e$ the reduced flux quantum and $\Phi$ the flux threading the ring~\cite{Bergeal2010,Abdo2013c}. Three wave mixing occurs by embedding the ring in resonant circuits (Fig.~\ref{fig1}(b)) so that each flux can be expressed as  $\varphi_k\propto(\hat{k}+\hat{k}^\dag)$ where $\hat{k}$ is the canonical annihilation operator of a microwave mode of characteristic impedance $Z_k$ and resonance frequency $f_k$.
Although the Josephson mixer can be used in various ways~\cite{Bergeal2010,Abdo2013c,Bergeal2010b,Roch2012,Flurin2012,Abdo2013,Flurin2013,Abdo2013b}, we will focus here on the amplification regime in order to describe the figures of merit on a concrete case. The signals sent towards $a$ and $b$ modes are amplified in reflection in a phase preserving manner by driving the mode $p$ out of resonance at the frequency $f_a+f_b$~\cite{Bergeal2010}. Three main specifications matter in analog processing of quantum microwave signals. First, the power gain $G$ of the amplifier needs to be large enough so that the quantum noise at the input of the amplifier dominates all other noise sources on the detection setup. Typically 20~dB is enough if a cryogenic HEMT is used as a second stage of amplification~\cite{Weinreb1988}. Second, the time correlations of the quantum signals should be dominated by the system of interest and not by the Josephson mixer. This requires to have as large a dynamical frequency bandwidth $\Gamma$ as possible. Finally, the maximum input power $P_{in}^\mathrm{max}$ that does not affect the gain by more than 1~dB needs to be large enough to avoid any limitation on the amplitude of the quantum signals.

Optimizing these three parameters for a practical amplifier has been at the center of recent experimental works in various geometries~\cite{Hover2012,Eom2012,Mutus2013,Mutus2014}. Recently degenerate Josephson Parametric Amplifiers have reached more than $15$~dB gain over a $700$~MHz bandwidth in an all lumped-element based design\cite{Mutus2014} while $15$~dB gain over GHz bandwidth has been reported in a TiN traveling wave parametric amplifiers \cite{Eom2012,Bockstiegel2014}. The constraints on the parameters of the Josephson mixer are similar in origin to those of its degenerate cousin, the Josephson Parametric Amplifier~\cite{Vlad2007,Eichler2014b}, but with some differences~\cite{Abdo2013c}. First, there is an upper bound on the energy that is stored in the $p$ pump mode, originating from the small flux $\varphi_p$ assumption in the three-wave mixing term $H_{mix}$. Therefore, in order to allow pump powers to reach the onset of parametric oscillations, at which large gain $G$ develops, one has to ensure that~\cite{Abdo2013c}
\begin{equation} 
p_a Q_a p_b Q_b> \Xi,
\label{pqeq}
\end{equation}
where  $\Xi$ is a number depending on the exact geometry~\cite{Vlad2007} of the mixer.  $\Xi=8$ will be used in the following\cite{Manu2014}. In this expression, $Q_k$ is the quality factor of mode $k$, defined as $Q_k=f_k/\Gamma_k$  where $\Gamma_k$ is the resonance bandwidth. The participation ratio $p_k$ of the JRM Josephson junctions in the $k$ mode quantifies the ratio of the total energy in this mode that is actually stored across the JRM junctions. Second, there is an upper bound on the power spectral density of the amplified signals coming  from the the small flux $\varphi_{a,b}$ assumption in the three-wave mixing term $H_{mix}$. Indeed, neither the input signal power $P_{in}$, nor the vacuum noise should be amplified beyond a fraction of the Josephson energy $E_J$. For large gain, this condition can be approximated as\begin{equation} 
p_kG\left(P_{in}/(2\pi\Gamma_k)+hf\right) \Xi'<E_J,
\label{qpeq}
\end{equation}
where $\Xi'$ is a number of order 1 and $E_J$ is the Josephson energy $E_J=\varphi_0^2/L_J$. The two above constraints (\ref{pqeq}) and (\ref{qpeq}) indicate that increasing both $P_{in}^\mathrm{max}$ and $\Gamma$ for a given gain $G$ requires to increase the participation ratios $p_k$ and the Josephson junction energy $E_J$. However, these two figures are related in general since the Josephson inductance of a single junction of the JRM decreases with $E_J$ as $L_J=\varphi_0^2/E_J$. An easy way to set $p_k=1$ whatever the value of $E_J$ is achieved when all the inductive parts of the resonators originate from the Josephson junction themselves (Fig.~\ref{fig1}(c)). 

In practice, spurious geometric inductances develop due to the finite size of the circuit. For an extra inductance $L^{k}_{ex}$ in series with the junction, one gets $p_k=L_{J}/({L_{J}+ L^{k}_{ex}})$~(Fig. \ref{fig1}(d)). It is enlightening to represent graphically the constraints (\ref{pqeq},\ref{qpeq}) in the parameter phase spaces. In figure \ref{fig1}e, shaded areas delimitate, for two different values of $L_{ex}$, the allowed values of the quality factor $Q$ and Josephson inductance $L_J$ for a typical quantum limited amplifier operating at a frequency $f=7~$GHz with a $20~$dB power gain, a saturating input power $P_{in}^{max}=-110~$dBm. Note that for the sake of clarity the two $a$ and $b$ modes have been set to identical parameters. As can be seen in the figure, lower $Q$ (larger bandwidth) can be obtained only by lowering $L_{ex}$. Conversely, Fig. \ref{fig1}(f) shows the maximal allowed extra inductance $L_{ex}^{max}$ as a function of saturating power $P_{in}^\mathrm{max}$ for several desired quality factors $Q$. From these curves, one can deduce which maximal extra inductance $L_{ex}^{max}$ can be used for a given bandwidth. If $L_{ex}$ comes from the geometrical inductance of some wires, their length is of the order of $L_{ex}/\mu_0$, which is represented on the right axis of Fig.~\ref{fig1}(f). From these considerations, one also determines the maximal spatial extension of a Josephson mixer to ensure a given bandwidth.

\begin{figure}
\begin{centering}
\includegraphics[scale=.9]{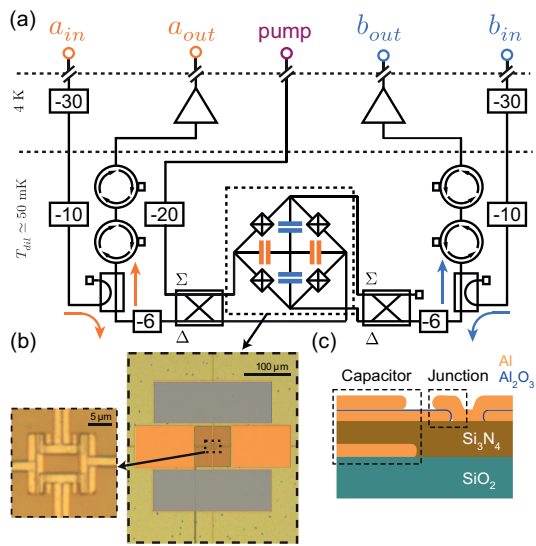}
\label{fig2}
\caption{\label{fig:setup} \textbf{(a)} Simplified schematic of the experimental setup. Differential $a$ and $b$ modes of the Josephson mixer are addressed in reflection through two 180$^o$ hybrid couplers. All input lines are filtered and attenuated (partially shown).
Output signals are separated from input signals by a directional
coupler and amplified by a low noise HEMT amplifier at 4K. \textbf{(b)} Optical microscope picture of the device showing the planar capacitors (right) and the Josephson junction ring (left).  \textbf{(c)} Side view of the device. The thickness of the SiO$_{2}$ is 300 nm, bottom plate of the capacitors is 35 nm and burried below 200 nm of silicon nitride, the top plate of the capacitors and the Josephson junctions are obtained by double angle deposition of 100 nm and 120 nm of aluminum with an intermediate oxydation.}
\end{centering}
\end{figure}

\begin{figure}
\begin{centering}
\includegraphics[scale=.65]{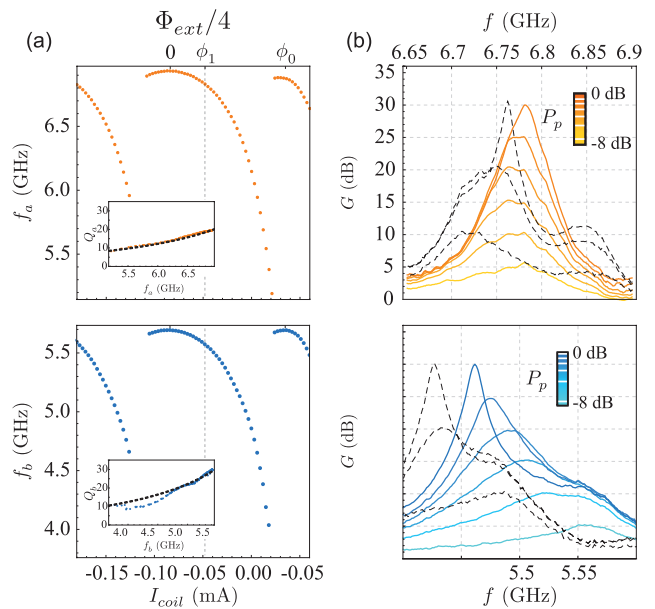}
\caption{\label{fig:flux dependence of the frequency}\textbf{(a)} Measured resonant frequencies of resonators $a$ and $b$ as a function of increasing current $I_\mathrm{coil}$ in the coil generating the flux bias $\Phi_\mathrm{ext}$ of the JRM when the pump is turned off. The insets show the measured quality factors $Q_k$ as a function of resonance frequency, where the dashed lines correspond to the predicted $Q_k$. The corresponding bandwidths for resonators $a$ and $b$  at the flux bias $\phi_{1}$ are respectively $\Gamma_a=365$~MHz. and $\Gamma_b=200$~MHz. \textbf{(b)} Gain in reflection
at the flux bias $\phi_{1}$ indicated as a line in \textbf{(a)}. The color bar encodes the pump power referred to the parametric oscillation threshold. The pump frequency is set to $12.26$~GHz. The black dashed lines show the gain of the amplifier obtained without the $6$~dB attenuator on the $\Delta$ ports of the hybrid couplers and during another cool down (hence the slightly different center frequency). The flux is close to $\phi_1$ and the pump frequency set to $12.26$~GHz.}
\end{centering}
\end{figure}

Our implementation of the design in (Fig.~\ref{fig1}(c)) is presented in Fig.~\ref{fig:setup}. The $a$ and $b$ mode resonators are composed of the JRM that is shunted by a cross of plate capacitors. The circuit is fabricated on a 500 $\mu$m thick Si chip covered with a 300 nm layer of SiO$_{2}$
on top. In a first step, a Ti(5nm)/Al(30nm) common counter electrode
for all the plate capacitors is fabricated using standard e-beam lithography (dark yellow in Fig.~\ref{fig:setup}(b)). It spreads all over the surface underneath the rest of the circuit, except for a hole in the center and a thin stripe (brown in Fig.~\ref{fig:setup}(b)) allowing to flux bias the circuit without the constraints imposed by the Meissner effect. The whole chip is then covered with 200 nm of amorphous dielectric silicon nitride by plasma-enhanced chemical vapor deposition. Finally,
the second metallic plate of the capacitors (area $285$~$\mu$m $\times$ $86$~$\mu$m for the $a$ resonator and $140$~$\mu$m $\times$ $86$~$\mu$m for the $b$ resonator) and the Josephson junctions  (area $4.2$~$\mu$m $\times$ $1$~$\mu$m) are fabricated by double angle deposition of 100 nm and 120 nm of aluminum with an intermediate oxydation step (Fig.~\ref{fig:setup}(c)). The circuit is then placed in a copper box enclosed in a Cryoperm
magnetic shielding box anchored at base temperature of a
dilution refrigerator ($T_{dil}\simeq50~$mK). 
Two 180$^o$ hybrid couplers address separately
the differential $a$ and $b$ modes through their $\Delta$ ports as
well as the pump mode $c$ through one of the $\Sigma$ ports. The resonators have characteristic impedances smaller than $10$~$\Omega$ and are galvanically connected to the $50$~$\Omega$ ports of the device so as to maximize bandwidth, only limited by impedance mismatch. Owing to the large coupling between the differential modes and the input/output ports, the gain is sensitive to the frequency dependence of the impedance~\cite{Mutus2014}. In order to probe the characteristics of the Josephson mixer alone without carefully engineering the impedance of the environment, we connect a $6$~dB attenuator on the $\Delta$ ports of the hybrid couplers. The impact of the $6$~dB attenuator can be seen on the spectral response of the mixer in Fig.~\ref{fig:flux dependence of the frequency}(b). A coil allows to control the flux threading the JRM loop to tune the mixing term in $H_{mix}$ via its current $I_\mathrm{coil}$.

The effect of the flux bias can also be seen (Fig.~\ref{fig:flux dependence of the frequency}(a)) on the resonance frequencies $f_a$ and $f_b$ which depend in an hysteretic manner of the flux with a period $4\phi_0$~\cite{Bergeal2010}. This hysteretic behavior could be removed by inserting additional inductances in the JRM in order to extend the static bandwidth of the amplifier~\cite{Roch2012}. However, this comes at the expense of lowering participation ratios, which we aim at maximizing, and becomes less useful with a large dynamical bandwidth. 
In this device, we observe a frequency dependence on $I_{coil}$, which is not perfectly periodic. This observed non-linear dependence of $\Phi_{ext}$ on $I_{coil}$ may originate form vortex dynamics in the large superconducting capacitor plate that is buried under the silicon nitride. For each resonance frequency, it was possible to measure the quality factor (inset of Fig.~\ref{fig:flux dependence of the frequency}(a)). This dependence can lead to a quantitative model describing the Josephson mixer. In this detailed model, based on Fig.~\ref{fig1}(d), a stray inductance $L_{stray}$ is considered in series with the capacitor $C_{res}$. Using first full 3D microwave simulations of the whole device, it was possible to estimate the geometrical electrical parameters of $C^a_{res}=3$~pF, $L^a_{stray}+L^a_{ex}=130$~pH and $C^b_{res}=6$~pF, $L^b_{stray}+L^b_{ex}=85$~pH. Then, by fitting $L_J=90~\mathrm{pH}$, one gets $f_a^{(fit)}=6.95~\mathrm{GHz}$ and $f_b^{(fit)}=5.7~\mathrm{GHz}$, which are close to the measured resonance frequencies at $\Phi_{ext}=0$ (Fig.~\ref{fig:flux dependence of the frequency}(a)). From there, one can estimate the participation ratios to be $p_a=25\%$ and $p_b=35\%$. Note that similar values for the participation ratios can be obtained by fitting directly the flux dependence of the resonance frequency~(see supplementary material of Ref.~\cite{Flurin2013}). Finally, one can fit the stray inductances to best recover the curves $Q(f)$ derived in \cite{suppmat} and find $L^a_{stray}=75$~pH, $L^b_{stray}=51$~pH. 
The Josephson mixer can be used as an amplifier by setting the flux $\phi_{1}$ slightly lower than $2\phi_0=h/e$ and driving the $c$ pump mode at the frequency $f_p=12.26$~GHz, which is close to $f_a+f_b$~\cite{Bergeal2010}. The gain, which is the ratio of the reflected power when the pump is on and off, was measured in reflection on both amplifier ports as a function of frequency (Fig.~\ref{fig:flux dependence of the frequency}(b)) for various values of the pump power $P_p$ at the fixed flux $\phi_{1}$. As the pump power rises towards the parametric oscillation threshold, the gain increases on both ports up to 30 dB. Conversely, the operating bandwidth decreases. This curve demonstrates a bandwidth of  $50~\mathrm{MHz}$ at 20~dB, which is an order of magnitude higher than using previous implementations of the Josephson mixer~\cite{Bergeal2010b, Abdo2011, Roch2012}. 

\begin{figure}
\begin{centering}
\includegraphics[scale=.4]{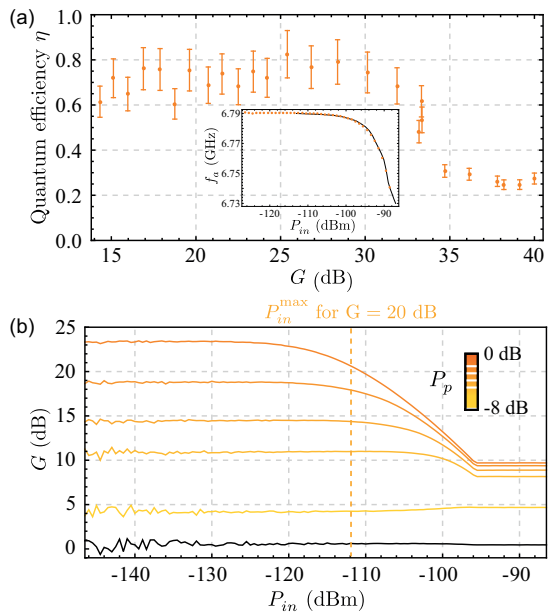}
\caption{\label{fig:Dynamic range}\textbf{(a)} Efficiency $\eta$ of the amplifier as a function of the gain measured on port $a$. The inset shows a comparison between experiments (orange points) and numerical simulation (continuous line) of the dependence of the resonant frequency $f_a$ on input power $P_{in}$. This allows an \emph{in situ} calibration of the input power at the level of the amplifier and thus deduce the attenuation of the input and output line connecting the amplifier to the instruments. The error bars on the efficiency corresponds to an error of $\pm 0.3$~dB on this calibration. \textbf{(b)} Measured gain $G$ on the $a$ port as a function of input power $P_{in}$ at
$6.76$~GHz for various pump powers (black line: pump is OFF, from yellow to red: increasing pump powers  -8,-5,-4,-3,-2~dB). The dashed vertical line indicates the 1dB compression point at 20 dB.}
\end{centering}
\end{figure}

The amplifier added noise is evaluated by amplification of zero point fluctuations in $a$ and $b$ modes. A spectrum analyzer measures the spectral power density coming from $a_{out}$ and $b_{out}$ as a function of $P_p$. No signal is sent into $a_{in}$ and $b_{in}$. The difference between spectral densities while the amplifier is ON and OFF is given by
\begin{equation} 
S_{ON}-S_{OFF}=G_{LNA}\times hfG (S_{add}+1),
\label{spectralnoise2}
\end{equation}
where $G$ is the gain of the amplifier and $G_{LNA}$ is the total gain of the output lines. Here the modes are assumed to be in the vacuum state and $G\gg 1$. In this case the added noise $S_{add}$ can be related to the quantum efficiency $\eta$ of the mixer\cite{Flurin2012} by $S_{add}=(1-\eta)/2\eta$. 
Determination of $\eta$ requires thus separate measurements of the amplifier gain $G$, as those of Fig. \ref{fig:flux dependence of the frequency}, and fine calibrations of the attenuations and gains of the input and output lines connecting the mixer to the detectors. We obtain this calibration by measuring the shift in frequency as a function of power sent into $a_{in}$ and $b_{in}$. It provides, by comparison with numerical calculations of the circuit\cite{suppmat}
 shown in Fig. \ref{fig1} (d), a precise calibration of the attenuation of the input line between $a_{in}$ and the mixer. The gain $G_{LNA}$ between the Josephson mixer and $a_{out}$ is then deduced from the total transmission between $a_{in}$ and $a_{out}$, the pump being turned off. Note that we also observe a clear frequency shift in the gain measurements of Fig. \ref{fig:flux dependence of the frequency} (b) while changing pump power due to higher order cross-Kerr~\cite{Hoi2013} terms that are proportional to $P_{p} a^{\dagger}a$ and $P_{p} b^{\dagger}b$.
Figure~\ref{fig:Dynamic range}(a) presents the measured $\eta$ as a function $G$. It indicates an efficiency of $0.7$, in agreement with an independent measurement\cite{suppmat}, up to $33$~dB of gain above which the amplifier enters the parametric oscillation regime where $\eta$ is near $0.2$.

The last important specification of an amplifier is its dynamical range. It is characterized by the 1 dB compression point $P_{in}^\mathrm{max}$ of the amplifier. In Josephson parametric devices, such as the Josephson mixer, this saturation can be caused either by depletion of the pump (\emph{i.e.} the gain is so large that the
pump cannot refill quickly enough to feed the amplifier), or
by reaching a large enough number of photons such that higher order non-linearities cannot be neglected. Using the calibration of the input power $P_{in}$ on port $b$ above, we used a vector network analyzer to measure the output power $P_{out}$ as a function of
input power $P_{in}$ (Fig. \ref{fig:Dynamic range}(b)) at $6.76$~GHz (center frequency when $G=20$~dB) for various pump powers following Ref.~\cite{Abdo2013c}. At low input powers $P_{in}<-120~\mathrm{dBm}$, the gain goes from 0 to 25~dB for increasing pump power $P_p$ without depending on $P_{in}$. For the pump power corresponding to $G=20$~dB at low input power, the amplifier behaves linearly for low power until it reaches the $1$~dB compression point at  $-112$~dBm at the JRM input. 
At $6.76$~GHz and for a bandwidth of $50$~MHz, this power corresponds to $4.5$ photons per bandwidth.
Above this threshold the gain drops and finally
saturates. This dynamical range is large enough not only for performing qubit readout~\cite{Abdo2011} but also for amplifying vacuum squeezed states~\cite{Flurin2012}. 

In conclusion, we have discussed an efficient and compact design for the Josephson mixer and applied these principles to demonstrate phase preserving quantum limited amplification. The resulting device operates with gains reaching $30$~dB within 0.4 photons of the quantum limit of noise and a saturation power of $-112$~dBm or equivalently $4.5$ photons per bandwidth, which is promising for analog information processing of quantum signals, directional amplification and on-chip circulators \cite{Sliwa2015}. These specifications do not hinder the dynamical bandwidth of the mixer, which reaches $50$~MHz at $G=20$~dB. Such device is suited for fast operation on superconducting qubit, which are necessary to the improvement of the efficiency of quantum feedback \cite{Vijay2012,Campagne2013}, multiplexing several qubits \cite{Chen2012} or more generally quantum error correction schemes.

\begin{acknowledgements}
We thank Michel Devoret for enlightening discussions. Nanofabrication has been made
within the consortium Salle Blanche Paris Centre. This work was supported by the program ANR-12-JCJC-TIQS of Agence Nationale pour la Recherche. JDP acknowledges financial support from Michel Devoret. \end{acknowledgements}

\end{document}